\documentstyle[preprint,aps]{revtex}

\begin{document}
\author{Lev I. Deych, D. Livdan, A.A. Lisyansky}
\title{Resonant tunneling of electromagnetic waves through polariton gaps}
\date{\today}
\draft
\maketitle
\tightenlines
\begin{abstract}
We consider resonance tunneling of electromagnetic waves through an optical
barrier formed by a stop-band between lower and upper polariton branches. We
show that the tunneling through this kind of barrier is qualitatively
different from tunneling through other optical barriers as well as from the
quantum mechanic tunneling through a rectangular barrier. We find that the
width of the resonance maximum of the transmission coefficient tends sharply
to zero as the frequency approaches the lower boundary of the stop band.
Resonance transmission peaks give rise to new photonic bands inside the
polariton stop band in a periodic array of the barriers.
\end{abstract}


\section{Introduction}

The effect of tunneling is well studied in the context of quantum mechanics
(see, for example, book \cite{Book}). Recently, tunneling of electromagnetic
waves has attracted interest owing to experiments with evanescent
electromagnetic waves, \cite{Enders,Steinberg,Ranfagni,Spielman} which are a
direct analog of wave functions of tunneling quantum particles. These
experiments provide an opportunity to experimentally study the time
evolution of tunneling wave packets -- a subject of long-standing
controversy (see, for example, review articles \cite
{Landauer,Hauge,Olkhovsky}). So far, three types of optical barriers have
been considered in the context of the tunneling experiments. Historically,
the first experiment with evanescent modes was carried out by Bose in 1927,
where a prism with a beam incident at an angle larger then the angle of
total internal reflection was used as a barrier \cite{Bose}. The same idea
was used by Yeh \cite{Yeh}, who considered resonant tunneling of
electromagnetic waves in a superlattice composed of alternating layers with
indexes of refraction such that the incident angle was greater then the
angle of total internal reflection for one layer and smaller for the other
one. The majority of time-of-flight theoretical considerations and
experiments have been performed for an undersized wave guide \cite
{Ranfagni,Enders,Martin}, and photonic band gaps.\cite{Steinberg,Spielman}
All of the considered optical barriers were artificial structures. However,
there exist natural optical barriers, which are well known, but were never
considered in the context of light tunneling. We are referring to
``restrahlen'' or ``stop'' photonic bands, which exist in many dielectrics
and semiconductors in the region of polariton resonances. It is well known
that in the absence of attenuation, the transmission coefficient of light
with a frequency within the stop band decreases exponentially with an
increase in the width of a slab through which it propagates. Therefore, a
simple slab of a dielectric with polariton resonances can serve as a
``naturally made'' photonic barrier. Properties of this polariton barrier,
however, might be considerably different from properties of other kinds of
barriers because of the different dispersion law of electromagnetic waves
tunneling through the polariton barrier. It is important, therefore, to
consider the tunneling properties of polariton barriers.

In this paper we consider tunneling of electromagnetic radiation through a
polariton barrier in a steady state mode. The point of interest in this
situation is the resonance tunneling through a two-barrier system. Resonance
tunneling of electromagnetic waves in a superlattice structure was
previously discussed in Ref.\cite{Yeh}. A tunneling effect in Ref.\cite{Yeh}
arose when the angle of incidence for electromagnetic waves exceeded the
angle of total internal reflection for one of the layers constituting the
superlattice. The frequency dependence of the imaginary wave number in this
case coincides with that of the square-barrier quantum tunneling problem.
The situation considered in this paper is considerably different. The
imaginary wave number in a polariton barrier layer depends upon the
frequency in a peculiar way, demonstrating a singularity near the lower
boundary, $\omega _{T}$, of the polariton gap. This dependence does not have
an analog in quantum mechanical systems. This leads to a new pattern of
resonance tunneling states.

Extending our system from two-barrier to a periodic array of alternating
transparent and barrier layers, we obtain a new kind of photon band {\em %
inside the initial forbidden band}. These bands arise from the tunneling
states and inherit many of their properties. In particular, we show that the
bandwidths of these bands are determined by the widths of the respective
resonances. The band structure of 2-D and 3-D photonic crystals made of
materials with frequency dependent dielectric permeability was numerically
studied in Ref.\cite{Maradudindisp1,Maradudindisp2,Zhang}. The authors of
Ref.\cite{Maradudindisp1} computed bands of two and three dimensional
structures composed of metal rods (spheres in the 3-D case) with dielectric
permeability of the form $\varepsilon =1-\omega _{p}^{2}/\omega ^{2}$, where 
$\omega _{p}$ is the frequency of plasma excitations in the metal. The
authors of Ref. \cite{Zhang} dealt with a photonic crystal built of a
dielectric with $\varepsilon $ similar to that used in the present paper
[see Eq.(\ref{epsilon}) below]. One of the interesting effects observed in
both studies was flattening of the photonic bands below the frequency $%
\omega _{p}$, in the case considered in Ref.\cite{Maradudindisp1}, or below $%
\omega _{T}$ for Ref.\cite{Zhang}. We show analytically that this effect in
the case of the dielectric photonic crystal is a direct consequence of the
singular behavior of the dielectric function at $\omega _{T}$. We also
discuss the influence of absorption upon tunneling properties of the
polariton barriers.

\section{Resonance tunneling in a three-layer system}

We consider propagation of electromagnetic radiation of frequency $\omega $
in a system that consists of two parallel absorbing dielectric layers, each
of thickness $a$, positioned along the $x$-axis and separated by a distance $%
b$. The dielectric permeability of the dielectric material is given by 
\begin{equation}
\varepsilon (\omega )=\varepsilon _{\infty }\frac{\left( \omega +i\gamma
\right) ^{2}-\omega _{L}^{2}}{\left( \omega +i\gamma \right) ^{2}-\omega
_{T}^{2}},  \label{epsilon}
\end{equation}
where $\gamma $ is the attenuation, and $\varepsilon _{\infty }$ is the
background dielectric permeability. Both layers are considered optically
isotropic, so that TE and TM waves are not coupled. This allows us to
neglect the vector nature of the electric field. We consider the case of
normal incidence so that the electric field in our system can be presented
as follows:

\begin{equation}
\begin{tabular}{lll}
$E(x)$ & $=e^{ik_{0}x}+re^{-ik_{0}x}$ & $0\geq x>-\infty ,$ \\ 
& $=a_{1}e^{ikx}+b_{1}e^{-ikx}$ & $a\geq x>0,\bigskip \medskip $ \\ 
& $=a_{2}e^{ik_{0}(x-a)}+b_{2}e^{-ik_{0}(x-a)}$ & $a+b\geq x>a,\bigskip
\medskip $ \\ 
& $=a_{3}e^{ik(x-a-b)}+b_{3}e^{-ik(x-a-b)}\qquad $ & $2a+b\geq x>a+b,$ \\ 
& $=te^{ik_{0}(x-2a-b)}$ & $x>2a+b,$%
\end{tabular}
\label{field}
\end{equation}
where $k_{0}=\omega /c$ is the wave-vector in vacuum, $c$ is the speed of
light in vacuum, and $r$, $a_{1}$, $b_{1}$, $a_{2}$, $b_{2}$, $a_{3}$, $%
b_{3} $ and $t$ are the complex amplitudes of the plane waves in each of the
five different regions. The wave number $k$ in the dielectric layers is
determined by the expression following from Eq.\thinspace (\ref{epsilon}): 
\begin{equation}
k=k_{0}\sqrt{\varepsilon _{\infty }\frac{\left( \omega +i\gamma \right)
^{2}-\omega _{L}^{2}}{\left( \omega +i\gamma \right) ^{2}-\omega _{T}^{2}}}.
\label{wavenumber}
\end{equation}

Taking into account regular boundary conditions at each of the boundaries
between different materials, we obtain the following expression for the
complex transmission coefficient $t$: 
\begin{eqnarray}
t &=&16\left( \frac{k}{k_{0}}\right) ^{2}e\left. ^{2ika+ik_{0}b}\left( 1-%
\frac{k^{2}}{k_{0}^{2}}\right) ^{-2}\right[
2e^{2ika+ik_{0}b}-2e^{2ika}-e^{ik_{0}b}-e^{4ika+ik_{0}b}+  \label{t} \\
&&\left( 1-4\frac{k}{k_{0}}+6\frac{k^{2}}{k_{0}^{2}}-4\frac{k^{3}}{k_{0}^{3}}%
+\frac{k^{4}}{k_{0}^{4}}\right) e^{4ika}+1+4\frac{k}{k_{0}}+6\frac{k^{2}}{%
k_{0}^{2}}+4\frac{k^{3}}{k_{0}^{3}}+\left. \frac{k^{4}}{k_{0}^{4}}\right]
^{-1}.  \nonumber
\end{eqnarray}
In the case of a lossless dielectric ($\gamma =0$) the expression for the
transmissivity of the system, $T=\mid t^{2}|,$ can be obtained as 
\begin{equation}
T=\frac{1}{1+\left( \frac{\kappa }{k_{0}}+\frac{k_{0}}{\kappa }\right)
^{2}\sinh ^{2}(\kappa a)\cos ^{2}\left[ K(a+b)\right] },  \label{T}
\end{equation}
where $K(a+b)$ is given by

\begin{equation}
\cos \left[ K(a+b)\right] =\cosh (\kappa a)\cos (k_{0}b)+\frac{1}{2}\left( 
\frac{\kappa }{k_{0}}-\frac{k_{0}}{\kappa }\right) \sinh (\kappa a)\sin
(k_{0}b).  \label{K}
\end{equation}
We assume here that the frequency of the wave falls into the region $\omega
_{T}<\omega <\omega _{L}$ so that the wave number $k=i\kappa $ is imaginary,
and propagation of the wave through the dielectric layers is evanescent.
Eq.\thinspace (\ref{T}) describes optical tunneling through a forbidden
``band.'' However, for a set of frequencies such that 
\begin{equation}
\cos \left[ K(a+b)\right] =0,  \label{resonance}
\end{equation}
the transmission coefficient, $T$, is equal to $1$, and tunneling through
the system becomes resonant. Though Eq.\thinspace (\ref{K}) is similar to
the respective equations of Ref.\cite{Yeh} and of the quantum mechanics
tunneling problem, the different frequency dependence of the wave number $k$
causes different behavior of the solutions to this equation.

In order to determine the number of resonance frequencies, it is convenient
to rewrite Eq.\thinspace (\ref{resonance}) as 
\begin{equation}
\frac{1}{2}\left( \frac{k_{0}}{\kappa }-\frac{\kappa }{k_{0}}\right) \tanh
(\kappa a)=\cot (k_{0}b).  \label{solution}
\end{equation}
If boundary frequencies $\omega _{L}$ and $\omega _{T}$ satisfy the
inequality 
\begin{equation}
\omega _{L}-\omega _{T}<\frac{\pi c}{b},  \label{inequality}
\end{equation}
the respective interval of wave numbers form $\omega _{T}/c$ to $\omega
_{L}/c$ can only accommodate less than one period of $\cot (k_{0}b)$. In
this case Eq.\thinspace (\ref{solution}) has one solution if 
\begin{equation}
k_{0}a\tan (k_{0}b)>2.  \label{onesolution}
\end{equation}
If inequality\thinspace (\ref{inequality}) does not hold, the number of
resonance frequencies is equal to 
\begin{equation}
N=\frac{(\omega _{L}-\omega _{T})b}{\pi c}+1  \label{N}
\end{equation}
if inequality\thinspace (\ref{onesolution}) is satisfied, and it is $N-1$
otherwise. The transmission, $T$, as a function of frequency is presented in
Figs.1 for two different values of the widths of the layers $a$ and $b$. One
can see that as the ratio of $a$ to $b$ increases, so does the number of
resonant peaks.

Both plots in Fig.\thinspace 1 show a similar pattern: the separation
between resonant peaks remains approximately constant while the half-width $%
\Gamma $ of peaks sharply decreases as the frequency approaches $\omega _{T}$%
. The resonant half-width, $\Gamma ,$ can be found by expanding the
denominator of Eq.\thinspace (\ref{T}) near the resonance frequency $\omega
_{r}$ up to the term quadratic in $\omega -\omega _{r}$. If a resonance
frequency, $\omega _{r},$ is close to the lower boundary of the gap, $\omega
_{T}$, one can find for $\Gamma $ 
\begin{equation}
\Gamma \approx \frac{2\sqrt{2}}{k_{0}a}\frac{(\delta \omega )^{2}}{\omega
_{\ast }}\exp {\left[ -2(k_{0}a)\left( \frac{\omega _{\ast }}{\delta \omega }%
\right) ^{1/2}\right] ,}  \label{Gammaleft}
\end{equation}
where $\delta \omega =\omega _{T}-\omega _{r}$, and $\omega _{\ast
}=\epsilon (\omega _{L}^{2}-\omega _{T}^{2})/(2\omega _{T})$. Eq. (\ref
{Gammaleft}) describes an extremely sharp non-analytical behavior of the
width of the resonance near the low-frequency boundary of the gap.

To take absorption into account, one can replace the parameter $\delta
\omega $ by the following expression 
\[
\delta \omega _{abs}=\sqrt{(\omega _{T}-\omega _{r})^{2}+\gamma ^{2}}. 
\]
If $\omega _{T}-\omega _{r}<\gamma $ the width of the resonance becomes
proportional to $\exp {(-\omega _{\ast }/\gamma )^{1/2}}$ and remains small
for $\gamma <\omega _{\ast }$. However, the width grows exponentially fast
with an increase of the absorption rate, and the height of the peak
decreases correspondingly. Therefore, the resonances nearest to $\omega _{T}$
are the most vulnerable with respect to absorption. One can see this in
Fig.2, which shows that these resonances disappear first when the absorption
becomes larger. It also follows from this analysis that the width of the
resonance peaks close to $\omega _{T}$ is determined solely by absorption.
This behavior of the resonances is specific to our particular model and does
not take place in the case of Ref. \cite{Yeh}, where the wave-vector is
finite for any $\omega $ in the forbidden band.

When $\omega _{r}$ approaches $\omega _{L}$, the wave-vector $\kappa (\omega
_{r})$ becomes smaller and $\Gamma $ increases with frequency, reaching at $%
\omega _{0}\simeq \omega _{L}$ a certain value, which depends upon all the
parameters of the system. For these frequencies, absorption, when small
enough, does not contribute significantly to the width of the maxima.

\section{Photonic bands in the polariton forbidden gap}

In the case of a periodic array of the three-layer sandwiches considered in
the previous section of the paper, one has a set of propagating bands
instead of single resonance frequencies. In this situation, Eq.\thinspace (%
\ref{K}) presents the dispersion equation of the propagating modes with $K$
being the Bloch wave number. The resonant tunneling frequencies discussed in
the previous section are the solutions of the dispersion equation at the
center of the Brillouin band, $K(a+b)=\pi /2$, where $K$ changes between $0$
and $\pi $ within the band. It is clear, therefore, that each of the
resonances gives rise to a corresponding band. This conclusion is supported
by the dispersion curves plotted in Figs. 3. One can see from this figure
that for each resonant frequency $\omega _{rn}$, where $n$ is the number of
a given mode, there is a distinct branch of the dispersion curve. It is
interesting to note that branches corresponding to the resonant frequencies
close to $\omega _{T}$, have much smaller dispersion than branches which
correspond to $\omega _{r}\simeq \omega _{L}$. One can also notice that
regular bands, which appear outside of the polariton gap, also become less
dispersive for frequencies near $\omega _{T}$. This observation agrees with
the results reported in Ref.\cite{Maradudindisp1,Maradudindisp2,Zhang} for
two-dimensional systems. The advantage of the one-dimensional model is that
we can analytically determine the cause for this behavior. Assuming that the
width of a band, centered at $\omega _{rn}$, is much smaller then the
frequency itself, one can expand the right-hand side of the dispersion
equation (\ref{K}) with respect to $\omega -\omega _{rn}$ and obtain the
approximate dispersion equation as follows: 
\begin{equation}  \label{dispersion}
\omega -\omega _{rn}=(-1)^{n}\Gamma \left( \omega _{0n}\right) \left( \frac{%
\kappa }{k_{0}}+\frac{k_{0}}{\kappa }\right) \sinh \left[ \kappa (\omega
_{0n})a\right] \cos \left[ K(a+b)\right] .  \label{dispersionappr}
\end{equation}
If one combines this equation and Eq.\thinspace (\ref{Gammaleft}) for the
resonance width $\Gamma $, the expression for the band width, $\Delta \omega
_{n}$, for bands near the lower edge $\omega _{T\text{ }}$ becomes

\[
\Delta \omega _{n}\sim \frac{(\delta \omega )^{2}}{\omega _{\ast }}. 
\]
This expression also holds for frequencies at the pass-band side of $\omega
_{T}$ and explains, therefore, the flattening of the photon dispersion
curves near $\omega _{T}$ observed in Ref.\cite
{Maradudindisp1,Maradudindisp2,Zhang}. Both the band width $\Delta \omega
_{n}$ and the width of the resonance $\Gamma $ tend to a certain finite
value when $\omega $ approaches $\omega _{L}$. The prefactor $(-1)^{n}$ in
Eq.\thinspace (\ref{dispersionappr}) shows that the branches have an
alternating sign of the dispersion: positive for the ``even'' branches and
negative for the ``odd'' ones. This feature is clearly seen in Fig.3.

\section{Conclusion}

We consider a polariton stop-band (restrahlen region) as an optical barrier
for tunneling of electromagnetic waves. We show that the peculiar frequency
dependence of the tunneling penetration length of electromagnetic waves
results in tunneling properties that are qualitatively different from those
of other optical barriers. We carry out a detailed study of the resonant
tunneling of electromagnetic waves through a three-layer sandwich, where a
dispersionless (vacuum) layer was placed between two identical layers with
dielectric permeability allowing for the polariton stop-band. It was found
that the number of resonance frequencies depends upon the frequency width of
the stop-band, $\omega _{L}-\omega _{T},$ and the spacial width, $b,$ of the
vacuum layer. This number increases with an increase of $\omega _{L}-\omega
_{T}$ and $b$. The latter dependence appears paradoxical, since it implies
that no matter how far away tunneling layers are from each other, they are
capable of providing conditions for resonant tunneling. The situation
becomes clear if one recollects that we deal with the steady state
situation, where the flux of energy in the system is fixed by an external
source. For a large system, a wave has to travel a long distance between the
layers before it reaches the steady state condition. This fact is crucial
for an experimental observation of resonance tunneling since one has to have
a steady enough source of light and be able to maintain a complete coherency
of the wave while it travels between the layers. Let us consider, for
example, $GaAs$, for which $\omega _{T}=5.1\times 10^{13}s^{-1}$, and $%
\omega _{L}=5.5\times 10^{13}s^{-1}$\cite{Kittel}. The vacuum wavelength in
this case changes between $34\mu m$ and $37\mu m$ when frequency sweeps over
the polariton gap. If one makes the distance $b$ between the layers equal
to, say, $350\mu m$ (approximately ten wavelengths), then the number of
observable resonances would be equal to $1$ or $2$ depending upon the width
of the dielectric layers. In order to observe greater number of the
resonance one could use materials with wider polariton gaps, for example, $%
MgO$, where $\omega _{T}=7.5\times 10^{13}s^{-1}$ and $\omega _{L}=14\times
10^{13}s^{-1}$\cite{Kittel}. In this case even for $b$ as small as $60\mu m$%
, four resonance peaks could be observed. The actual number of resonance
maxima also depends upon relaxation characteristics of the layers. Though
these characteristics can vary even for different samples of the same type
of material, one can consider $\gamma =0.02$ \cite{book} as a typical value
and use it for quantitative estimates. It is seen from Fig.2 that the
relaxation of this order of magnitude could reduce the number of observable
resonances by at most one maximum if the configuration of the system allows
for a maximum near $\omega _{T}$. More specific conclusions regarding the
number of the expected resonances could be drawn from the results of the
paper for each particular experimental configuration.

The width of the resonance peaks of the transmission coefficient was found
to have a sharp nonanalytical frequency dependence at the lower boundary of
the polariton gap, $\omega _{T}$ and to saturate at a certain finite value
when approaching the upper boundary, $\omega _{L}$. This fact results in
different reactions to absorption for resonances occurring in vicinities of $%
\omega _{T}$ and $\omega _{L}$ . In the first situation even a small
absorption completely determines the real width of the resonance, and these
resonances are washed out first with an increase of absorption. The
resonances in the latter case survive much stronger rate of absorption, and
for a small one they have a width mainly determined by the properties of the
barrier itself.

We also consider extension of our system to a periodic arrays of the
sandwiches. We show that each of the resonances of the original system gives
rise to a band, with the original resonance frequency at the center of the
respective band. The band widths of the bands (and correspondingly their
degrees of dispersion) are found to be determined by the frequency width of
the respective parent resonances. This fact explains flattening photonic
bands near $\omega _{T}$ observed previously in numerical simulations of 2-D
systems\cite{Zhang}.

We wish to thank S.A. Schwarz for reading and commenting on the manuscript.
This work was supported by the NSF under grant No. DMR-9632789, by a CUNY
collaborative grant, and by a PSC-CUNY research award.

\section*{CAPTIONS}

{\bf Figure 1}. Transmission as a function of $\frac{\omega }{\omega _{T}}$.
The ratio of $\frac{\omega _{L}}{\omega _{T}}$ is equal to $3$ for both
plots. (a) Values of $a$ and $b$ are $2$ and $8$ respectively. (b) $a=2$ and 
$b=20$. The polariton gap lies between normalized frequencies $1$ and $3$.

{\bf Figure 2}. Transmission in the presence of absorption for $a=2$ and $%
b=8 $. The dashed line corresponds to an attenuation coefficient $\frac{%
\gamma }{\omega _{T}}=0.005$, and the solid line depicts the transmission
for $\frac{\gamma }{\omega _{T}}=0.05$.

{\bf Figure 3}. The bands arising in the polariton gap in the case of
periodic arrangement of the considered system. (a) $a=2$, $b=8$, (b) $a=2$, $%
b=20$. The polariton gap is shown by dashed lines. Dispersion curves in the
forbidden gap correspond to the resonant peaks from Fig.~(1).

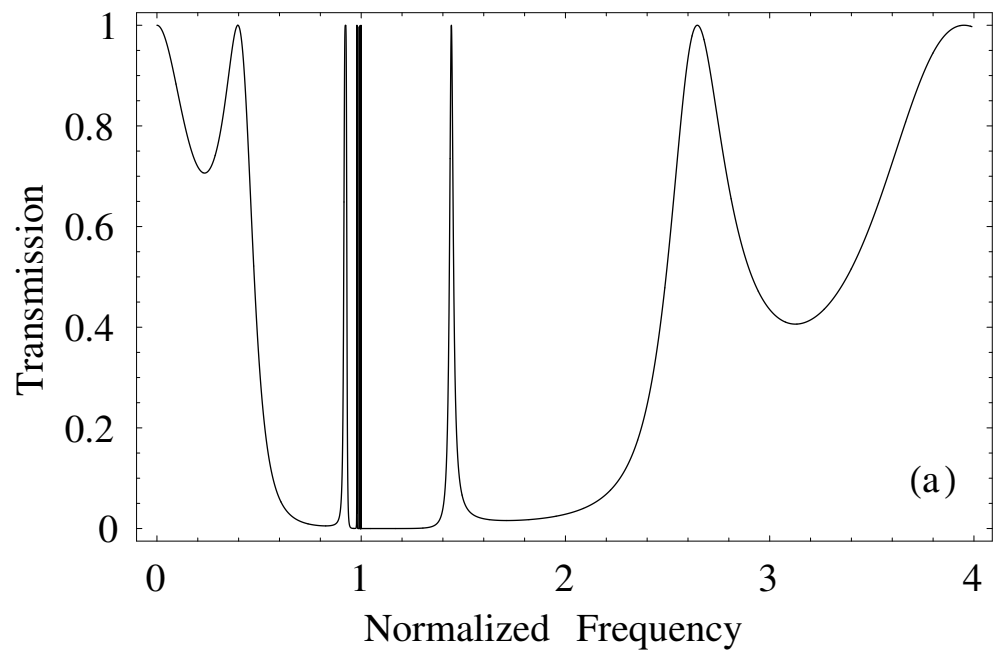
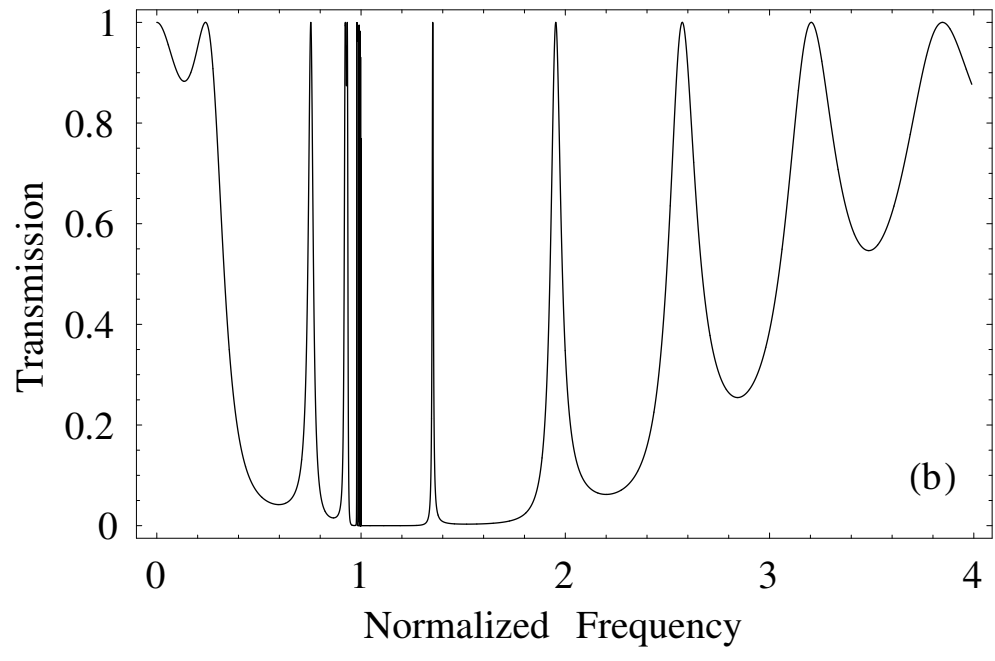
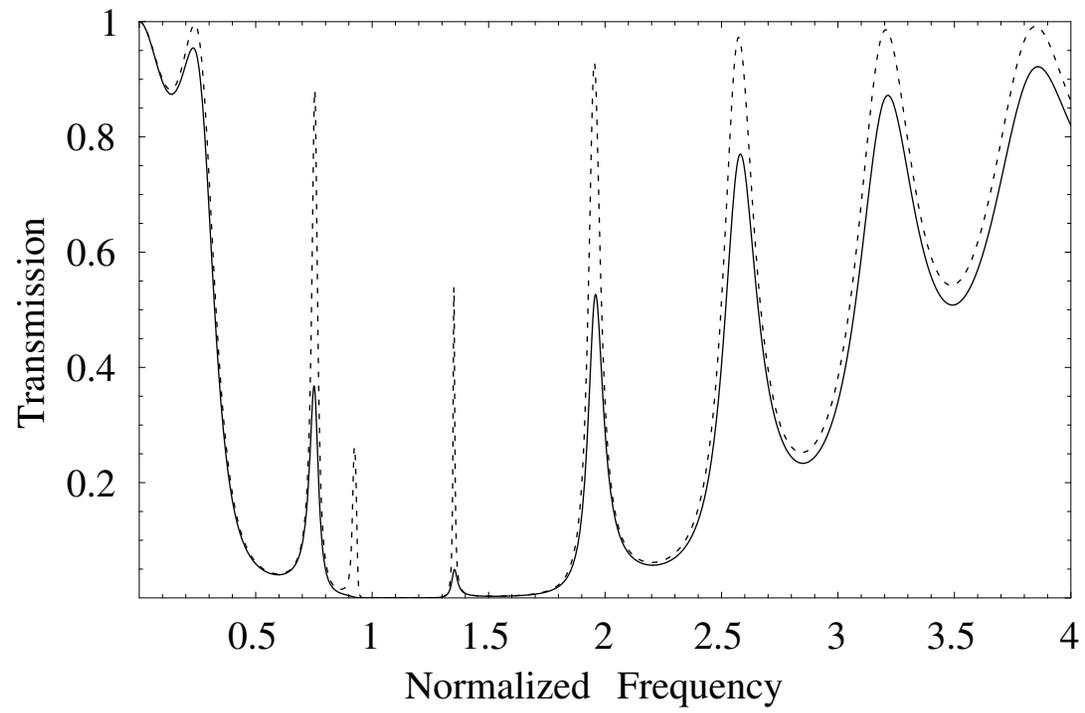
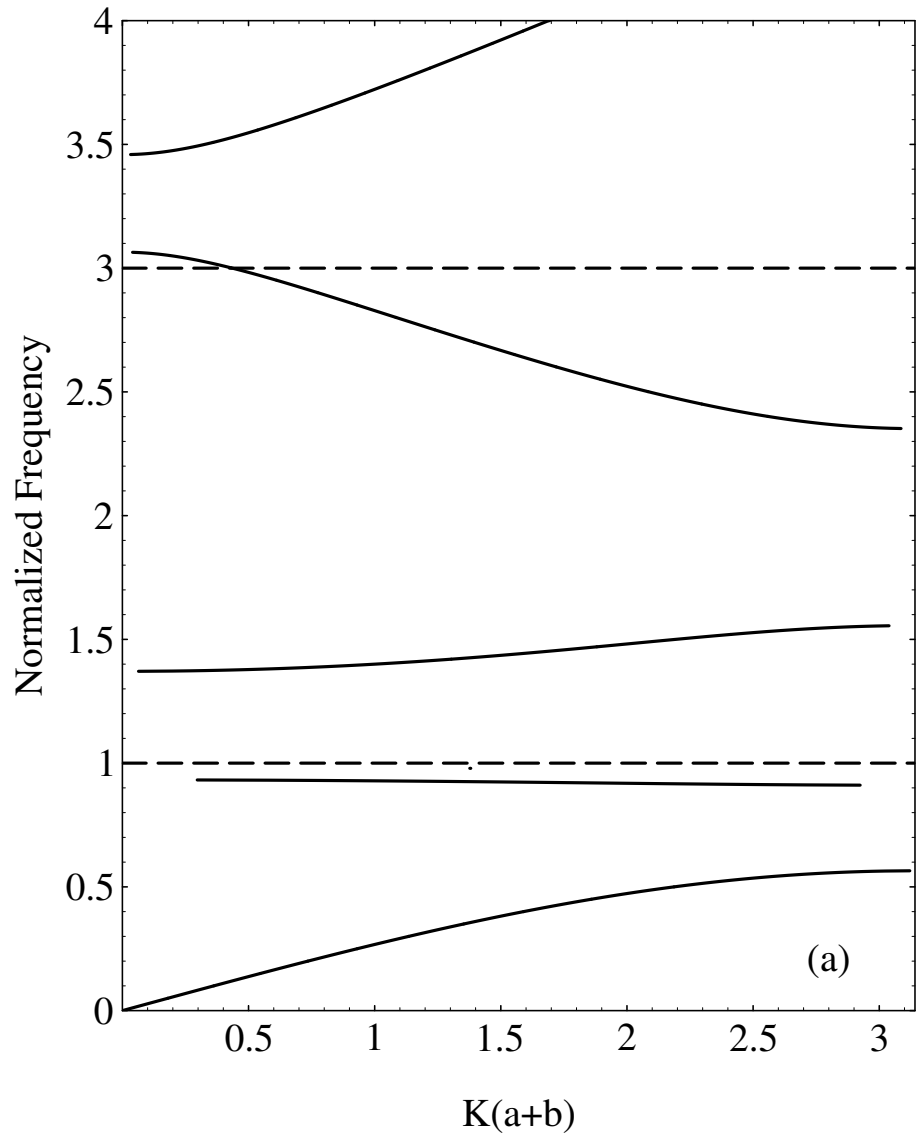
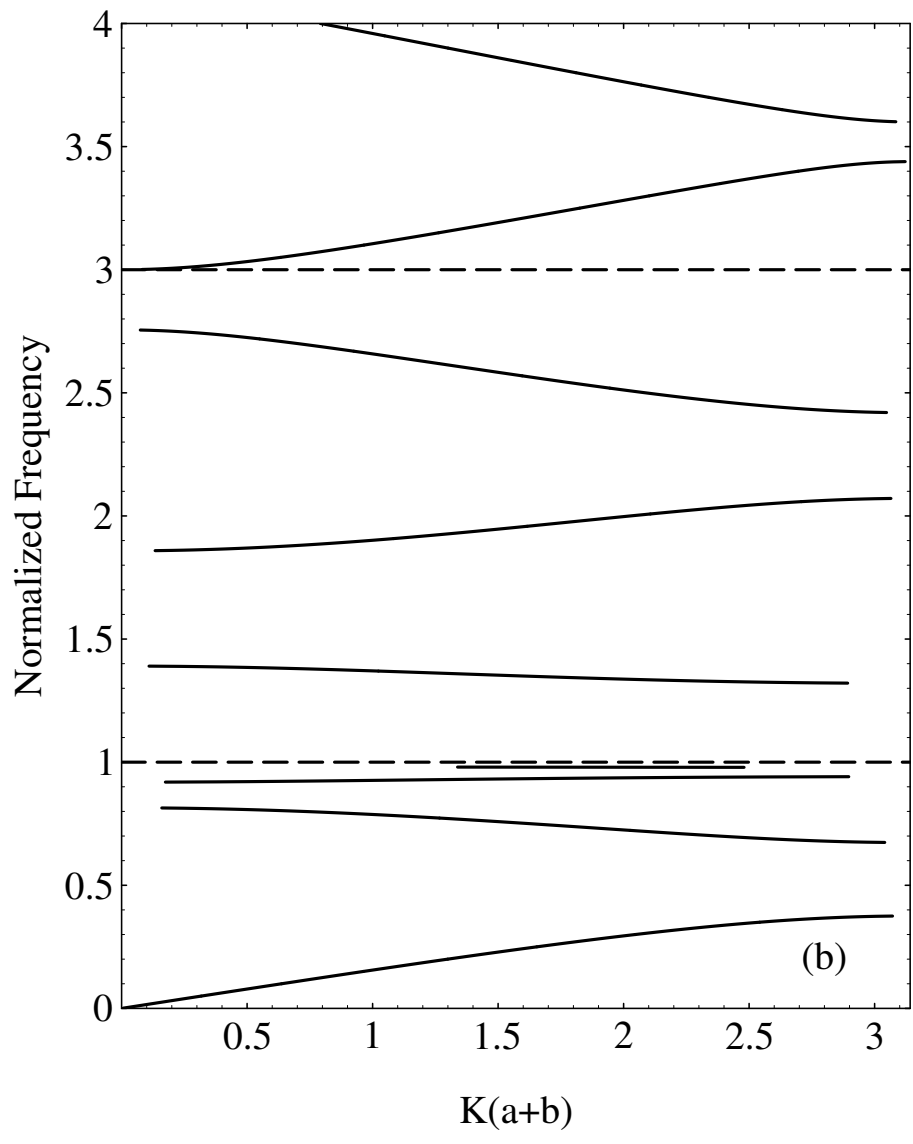
\end{document}